\documentclass[12pt]{article}
\usepackage{amsfonts,amssymb}
\usepackage{color}
\usepackage{epic}
\usepackage{graphics}

\def\lddots{\mathinner{\mkern1mu\raise1pt\hbox{.}\mkern2mu
\raise4pt\hbox{.}\mkern2mu\raise7pt\vbox{\kern7pt\hbox{.}}\mkern1mu}}
\makeatletter
\def\numberbysection{\@addtoreset{equation}{section}
\def\theequation{\thesection.\arabic{equation}}}
\makeatother

\numberbysection

\newcommand{\be}{\begin{eqnarray}}
\newcommand{\ee}{\end{eqnarray}}
\newcommand{\non}{\nonumber}

\begin{document}

\begin{titlepage}
\vskip 0.4cm
\strut\hfill
\vskip 0.8cm
\begin{center}

%\begin{center}

{\bf {\Large Systematic derivation of boundary Lax pairs}}\footnote{Proceedings contribution RAQIS'07, 11-14 September 2007, Annecy-Le-Vieux, France \\ (talk given by A.D., based on arXiv:0710.1538)}

\vspace{10mm}

{\large Jean Avan\footnote{avan@ptm.u-cergy.fr}$^{a}$ and Anastasia Doikou\footnote{doikou@bo.infn.it}$^{b}$}

\vspace{10mm}

{\small $^a$ LPTM, Universite de Cergy-Pontoise (CNRS UMR 8089), Saint-Martin 2\\
2 avenue Adolphe Chauvin, F-95302 Cergy-Pontoise Cedex, France}

{\small $^b$ University of Bologna, Physics Department, INFN Section \\
Via Irnerio 46, Bologna 40126, Italy}

\end{center}

\vfill

\begin{abstract}

We systematically derive the Lax
pair formulation for both discrete and continuum integrable classical
theories with consistent boundary conditions.

\end{abstract}

\vfill
\baselineskip=16pt

\end{titlepage}

\section{The discrete case}

Quadratic Poisson structures first appeared as the well-known
Sklyanin bracket \cite{sklb}. A more general form, characterized
by a pair of respectively skew symmetric and symmetric matrices $(r,\ s)$
appeared in \cite{maillet} in the formulation of
consistent Poisson structures
for non-ultralocal classical integrable field theories.
Finally it was shown \cite{lipar} that this was the natural quadratic form
a la Sklyanin for a non-skew-symmetric $r$-matrix, reading:
\be \Big \{L_1,\ L_2 \Big \} = \Big [r- r^{\pi},\ L_1 L_2 \Big ]
+L_1 (r+r^{\pi})L_2 - L_2 (r+r^{\pi})L_1.\ee
A typical situation when one considers naturally a quadratic
Poisson structure for the Lax matrix occurs when considering
discrete or continuous
integrable systems where the Lax matrix
depends on either a discrete or a continuous variable; the Lax
pair is thus associated to a point on the space-like lattice or
continuous line \cite{ft, kulishsklyanin}. Let us first examine
the discrete case where one considers a finite set of Lax
matrices $L_n$ labelled by $n \in {\mathbb N}$.

Lax representation of classical dynamical evolution equations
\cite{lax} is one key ingredient in the modern theory of classical
integrable systems \cite{GGKM}--\cite{BBT} together with
the associated notion of classical $r$-matrix \cite{skl, sts}.
Introduce the Lax pair ($L,\ A$) for discrete integrable
models \cite{abla} (see also  \cite{mcwoo} for statistical
systems), and the associated auxiliary problem (see e.g.
\cite{ft}) \be  && \psi_{n+1} = L_n\ \psi_n \non\\
&& \dot{\psi}_n = A_n\ \psi_n.  \label{lat} \ee From the latter
equations one may immediately obtain the discrete zero curvature condition:
\be \dot{L}_n = A_{n+1}\ L_n - L_n\ A_n.
\label{zero}\ee The monodromy matrix arises from the first equation
(\ref{lat}) (see e.g. \cite{ft}) \be T_{a}(\lambda) = L_{aN}(\lambda)
\ldots L_{a1}(\lambda) \label{trans0}\ee where index $a$
denotes the auxiliary space, and the indices $1, \ldots , N$
denote the sites of the one dimensional classical discrete model.

Consider now a skew symmetric classical $r$-matrix which is a
solution of the classical Yang-Baxter equation \cite{skl, sts}\be
\Big [r_{12}(\lambda_1-\lambda_2),\
r_{13}(\lambda_1)+r_{23}(\lambda_2) \Big ]+ \Big [
r_{13}(\lambda_1),\ r_{23}(\lambda_2) \Big ] =0, \label{clyb} \ee
and let $L$ satisfy the associated Sklyanin bracket \be \Big \{ L_{a}(\lambda),\ L_b(\mu)
\Big \} = \Big [ r_{ab}(\lambda-\mu),\ L_{a}(\lambda) L_b(\mu)
\Big ]. \label{clalg} \ee It is then immediate that (\ref{trans0}) also satisfies (\ref{clalg}). Use of
the latter equation shows that the quantities $tr
T(\lambda)^n$ provide charges in involution, that is \be \Big \{
tr\ T^{n}(\lambda),\ tr\ T^m(\mu) \Big \}=0 \ee which again is trivial
by virtue of (\ref{clalg}). In the simple $sl_2$ case the only non
trivial quantity is $tr T(\lambda) =t(\lambda)$, that is the usual
``bulk'' transfer matrix. In the bulk case in particular the zero curvature condition
(\ref{zero}) is realized by $L_n$ and \cite{ft}:
\be A_n(\lambda, \mu) =
t^{-1}(\lambda)\ tr_{a} \{ T_a(N,n;\lambda)\ r_{ab}(\lambda -\mu)\
T_a(n-1,1;\lambda) \} \label{laf} \ee  where we define
\be T_{a}(n,m;\lambda)= L_{an}(\lambda) L_{a
n-1}(\lambda) \ldots L_{am}(\lambda),  ~~~~n>m. \label{deff} \ee

We now generalize the procedure described in \cite{ft} for periodic boundary conditions
to the case of generic integrable ``boundary conditions''.
We propose a construction of
two types of monodromy and transfer
matrices, and associated Lax-type evolution equations, albeit incorporating
a supplementary set of non-dynamical parameters encapsulated
into a ``reflection'' matrix $K(\lambda)$. Any physical interpretation of the
$K$-matrix as a description of the ``boundary properties'' may not be appropriate
in all cases. We
should stress that this is the first time to our knowledge (see also \cite{avandoikou}) that
such an investigation is systematically undertaken. There are several related
studies regarding particular examples of open spin chains \cite{guan, lisa},
however the derivation of the corresponding Lax pair is
restricted to the Hamiltonian only and not to all associated
integrals of motion. In this study we present a
generic description independent of the choice of model, and we
derive the Lax pair for each one of the entailed boundary
integrals of motion.

Subsequently we shall deal with two types of classical
algebras, which are derived from
two known types of consistent quantum boundary conditions.
These boundary conditions are known as soliton preserving (SP), (see e.g.
\cite{sklyanin}--\cite{masa}), and soliton
non-preserving (SNP) \cite{durham, gand, dema}.
SNP boundary conditions have been also introduced
and studied for integrable quantum lattice systems
\cite{doikousnp}--\cite{crdo}. From the algebraic perspective the
two types of boundary conditions are associated with two distinct
algebras, i.e. the reflection algebra \cite{sklyanin}
and the twisted Yangian respectively \cite{molev, moras} (see also relevant studies 
\cite{dema, doikouy, crdo, mac, doikounp}).
It will be convenient for our purposes here to introduce some useful
notation: \be && \hat r_{ab}(\lambda) = r_{ba}(\lambda)
~~~\mbox{for SP}, ~~~~\hat r_{ab}(\lambda) = r_{ba}^{t_a
t_b}(\lambda) ~~~\mbox{for SNP} \non\\ && r^*_{ab}(\lambda)
=r_{ab}(\lambda) ~~~\mbox{for SP}, ~~~~ r^*_{ab}(\lambda) =
r_{ba}^{t_b}(-\lambda) ~~~\mbox{for SNP} \non\\ && \hat
r^*_{ab}(\lambda) =r_{ba}(\lambda) ~~~\mbox{for SP}, ~~~~\hat
r^*_{ab}(\lambda) =
r_{ab}^{t_a}(-\lambda) ~~~\mbox{for SNP} \label{notation00} \ee
The two types of monodromy matrices will respectively represent
the classical version of the reflection algebra ${\mathbb R}$,
and the twisted Yangian ${\mathbb T}$ written in the compact form:
(see e.g. \cite{sklyanin, maillet}): \be && \Big \{{\cal T}_1(\lambda_1),\
{\cal T}_2(\lambda_2) \Big \} = r_{12}(\lambda_1-\lambda_2){\cal
T}_{1}(\lambda_1){\cal T}_2(\lambda_2) -{\cal T}_1(\lambda_1)
{\cal T}_2(\lambda_2) \hat r_{12}(\lambda_1 -\lambda_2) \non\\ &&
+ {\cal T}_{1}(\lambda_1) \hat r^*_{12}(\lambda_1+\lambda_2){\cal
T}_2(\lambda_2)- {\cal T}_{2}(\lambda_2)
r^*_{12}(\lambda_1+\lambda_2){\cal T}_1(\lambda_1) \label{refc}
\ee where $\hat r,\ r^*,\ \hat r^*$ are defined in (\ref{notation00}).
In most cases, such as the $A^{(1)}_{{\cal N}
-1}$ $r$-matrices $r_{12}^{t_1 t_2} = r_{21}$ implying that in the
SNP case $r^*_{ab} = \hat r^*_{ab}$. In the case of the Yangian
$r$-matrix $r_{12} =r_{21}$, hence all the expressions above may
be written in a more symmetric form.

In order to construct representations of (\ref{refc}) yielding
a generating function of integrals of
motion one now introduces $c$-number (non-dynamical) representations
satisfying the purely algebraic condition (\ref{refc}) since:
\be \Big \{
K_1^{\pm}(\lambda_1),\ K_2^{\pm}(\lambda_2) \Big \}=0. \label{kso}
\ee
Taking now as $T(\lambda)$ any bulk monodromy matrix (\ref{trans0})
built from local $L$ matrices obeying
(\ref{clalg}) and defining in addition \be \hat T(\lambda) =
T^{-1}(-\lambda) ~~~\mbox{for SP}, ~~~~~\hat T(\lambda) =
T^{t}(-\lambda) ~~~\mbox{for SNP}.
\label{notation} \ee one shows that representations of the corresponding
algebras ${\mathbb R},\
{\mathbb T}$, are given by the following expression see e.g.
\cite{sklyanin, paper}: \be && {\cal T}(\lambda) = T(\lambda)\
K^{-}(\lambda)\ \hat T(\lambda). \label{reps} \ee For a detailed
proof see e.g. \cite{paper}.

Define now as generating function of the involutive quantities \be
t(\lambda)= tr\{ K^{+}(\lambda)\ {\cal T}(\lambda)\}. \label{gen}
\ee Due to (\ref{refc}) it is shown that \cite{sklyanin,
paper} \be  \Big \{t(\lambda_1),\ t(\lambda_2) \Big \} =0, ~~~
\lambda_1,\ \lambda_2 \in {\mathbb C}. \label{bint} \ee

Usually one considers the quantity $\ln\ t(\lambda)$ to get
{\it local} integrals of motion, however for the examples we are
going to examine here the expansion of $t(\lambda)$ is enough to
provide the associated local quantities as will be transparent in
the subsequent section. Finally one shows that time evolution of the local Lax matrix $L_n$
under generating Hamiltonian action of $t(\lambda)$ is given by:
\be {\dot L}_n(\mu) = {\mathbb A}_{n+1}(\lambda, \mu)\
L_n(\mu) - {\mathbb A}_n(\lambda, \mu)\ L_n(\mu), \ee where
${\mathbb A}_n$ is the modified (boundary) quantity,
\be {\mathbb A}_n(\lambda, \mu) &=& tr_a \Big (K_a^+(\lambda)\
T_a(N, n;\lambda)\ r_{ab}(\lambda-\mu)\ T_a(n-1, 1;\lambda)\
K^-_a(\lambda)\ \hat T_a(\lambda) \non\\ &+& K_a^+(\lambda)\
T_a(\lambda)\ K^-_a(\lambda)\ \hat T_a(1,n-1;\lambda)\ \hat
r_{ab}^*(\lambda +\mu)\ \hat T_a(n, N;\lambda) \Big )
\label{aan1} \ee where $T(n, m; \lambda)$ is defined in (\ref{deff}) and \be \hat T(m,n ;\lambda) =
\hat L_{am}(\lambda) \ldots \hat L_{an}(\lambda) ~~~~n>m. \ee
To prove (\ref{aan1}) we need
in addition to (\ref{clalg}) one more fundamental relation i.e.
\be \Big \{\hat L_{a}(\lambda), L_b(\mu) \Big \} = \hat
L_a(\lambda) \hat r^*_{ab}(\lambda)L_b(\mu) - L_b(\mu) \hat
r^*_{ab}(\lambda+\mu) \hat L_a(\lambda). \label{funda2} \ee Taking
into account (\ref{clalg}) and  the latter expressions we derive: \be
\Big \{t(\lambda),\ L_{bn}(\mu) \Big\}&=& tr_a \Big
(K_a^+(\lambda)\ T_a(N, n+1;\lambda)\  r_{ab}(\lambda-\mu)\ T_a(n,
1;\lambda)\ K^-_a(\lambda)\ \hat T(\lambda) \non\\ &+&
K_a^+(\lambda)\ T_a(\lambda)\ K^-_a(\lambda) \hat
T_a(1,n;\lambda)\ \hat r_{ab}^*(\lambda +\mu)\ \hat T_a(n+1,
N;\lambda) \Big  ) L_{bn}(\mu) \non\\ &-& L_{bn}(\mu)\ tr_a \Big (
K_a^+(\lambda)\ T_a(N,n;\lambda)\ r_{ab}(\lambda -\mu)\
T_a(n-1,1;\lambda)\ K^-_a(\lambda) \hat T(\lambda) \non\\ &+&
K_a^+(\lambda)\ T_a(\lambda)\ K_a^-(\lambda)\ \hat
T_a(1,n-1;\lambda)\ \hat r_{ab}^*(\lambda+\mu)\ \hat T_{a}(n,
N;\lambda) \Big ). \label{modan} \ee Expression (\ref{aan1})
is readily extracted from (\ref{modan}).

Special care should be taken at the boundary
points $n=1$ and $n=N+1$. Indeed we set: $T(N, N+1, \lambda)= T(0,1,\lambda) = \hat
T(1,0,\lambda) =\hat T(N+1, N, \lambda) ={\mathbb I}$.
We should stress that the derivation of the boundary Lax pair is universal,
namely the expressions (\ref{aan1}) are generic and independent
of the choice of $L,\ r$.
Note that a different construction of representations
of (\ref{refc}) was already given
in a very general setting in \cite{mailfrei2}. It is related to the formulation of non-ultralocal
integrable field theories on a lattice and extends the analysis of \cite{maillet}.

\subsection{Example}

We shall now examine a simple example, i.e. the open
generalized DST model, which may be seen as a lattice version of
the generalized (vector) NLS model, (see also
\cite{nls1, nls2, kundubas, ragn, paper}) for further details). The open Toda chain will also be
discussed as a limit of the DST model. We shall explicitly
evaluate the ``boundary'' Lax pairs for the first integrals of
motion. We focus here on the special case of the simplest rational non-dynamical
$r$-matrices \cite{young}
\be r(\lambda) = {{\mathbb P} \over \lambda}
~~~~\mbox{where} ~~~~{\mathbb P}=\sum_{i,j=1}^{N} E_{ij} \otimes
E_{ji} \label{rr} \ee ${\mathbb P}$ is the permutation operator,
and $~(E_{ij})_{kl} = \delta_{ik} \delta_{jl}$.

The Lax operator of the $gl({\cal N})$ DST model has the following
form: \be L(\lambda) = (\lambda - \sum_{j=1}^{{\cal N}-1} x^{(j)}
X^{(j)}) E_{11} +b\sum_{j=2}^{{\cal N}} E_{jj} +b\sum_{j=2}^{{\cal
N}}x^{(j-1)} E_{1j} - \sum_{j=2}^{{\cal N}}X^{(j-1)}E_{j1}  \ee
with $x^{(j)}_n,\ X^{(j)}_n$ being canonical variables. In \cite{paper} the first
non-trivial integral of motion for the SNP case, choosing the simplest
consistent value $K^{\pm} = {\mathbb I}$ was explicitly computed:
\be && {\cal H} = -{1\over 2} \sum_{n=1}^N {\mathbb
N}_n^2 -b \sum_{n=1}^N \sum_{j=1}^{{\cal N}-1}
X_n^{(j)}x_{n+1}^{(j)} - {1\over 2}\sum_{j=1}^{{\cal
N}-1}(X_N^{(j)}X_N^{(j)} +b^2x_1^{(j)} ) \non\\ && \mbox{where}
~~~{\mathbb N}_n = \sum_{j=1}^{{\cal N}-1}x_n^{(j)}X_{n}^{(j)}.
\label{ham0}\ee Our aim is now to determine the modified Lax
pair induced by the non-trivial integrable boundary conditions. We
shall focus here on the case of SNP boundary conditions, basically
because in the particular example we consider here such boundary
conditions are technically easier to study. Taking into account (\ref{aan1}) we
explicitly derive the modified Lax pair for the generalized
DST model with SNP boundary conditions. Indeed, after expanding
(\ref{aan1}) in powers of $\lambda^{-1}$, and recalling (\ref{rr}) we obtain the quantity
associated to the Hamiltonian (\ref{ham0}) \be && {\mathbb A}_n^{(2)} = \lambda E_{11} -
\sum_{j\neq 1} X_{n-1}^{(j-1)}E_{j1}  +b \sum_{j\neq 1}
x_{n}^{(j-1)}E_{1j}, ~~~n \in \{ 2, \ldots N\} \non\\ &&  {\mathbb
A}_1^{(2)}= \lambda E_{11} - b \sum_{j \neq 1} x_{1}^{(j-1)}E_{j1}
+b \sum_{j\neq 1}  x_{1}^{(j-1)}E_{1j}, \non\\ &&  {\mathbb
A}_{N+1}^{(2)} = \lambda E_{11} -  \sum_{j \neq 1}
X_{N}^{(j-1)}E_{j1}  + \sum_{j\neq 1}  X_{N}^{(j-1)}E_{1j}.
\label{aa} \ee It is worth stressing that in the $sl_2$ case
the SP and SNP boundary conditions coincide given that \be L^{-1}(-\lambda) =
V\ L^t(-\lambda)\ V, ~~~~V =\mbox{antid}(1, \ldots,1).
\label{gag} \ee The equations
of motion associated to the Hamiltonian (\ref{ham0}) may be readily
extracted by virtue of \be {\dot L}
= \Big \{ {\cal H}^{(2)},\ L \Big \}. \label{com}\ee
Alternatively the equations of motion may be
derived from the zero curvature condition, which the modified Lax pair
satisfies. It is clear that to each one of the higher local
charges a different quantify ${\mathbb A}_{n}^{(i)}$ is associated. Both
equations (\ref{com}), (\ref{zero}) lead naturally to the same
equations of motion, which for example in the $sl_2$ case read as: \be
&& {\dot x}_n = x_n^2 X_n +b  x_{n+1}, ~~~~{\dot X}_n = -x_n
X_n^{2} -b X_{n-1}, ~~~n\in \{2, \ldots N-1 \} \non\\  && {\dot
x}_1 = x_1^2 X_1 + b x_2, ~~~~{\dot X}_1 =  -x_1 X_1^2 - b x_1
\non\\ && {\dot x}_N = x_N^2 X_N +X_N, ~~~~ {\dot X}_N = -x_N
X_N^2 - bX_{N-1}. \ee
Note that the Toda model \cite{toda} may be seen as an appropriate limit of the $sl_2$ DST
model (see also \cite{skld}), and the corresponding boundary Hamiltonian, Lax pair and equations
of motion are easily obtained from the expressions above (for more details see \cite{paper}).

\section{The continuous case}

Let us now recall the basic notions regarding the Lax pair and the
zero curvature condition for a continuous integrable model following
essentially \cite{ft}. Define $\Psi$ as being a solution of the following set
of equations (see e.g. \cite{ft}) \be &&{\partial \Psi \over \partial
x} = {\mathbb U}(x,t, \lambda)  \Psi\label{dif1}\\ && {\partial \Psi
\over \partial t } = {\mathbb V}(x,t,\lambda) \Psi \label{dif2} \ee 
${\mathbb U},\ {\mathbb V}$ being in general $n \times n$ matrices with entries
defined as functions of complex valued dynamical fields, their derivatives,
and the spectral parameter $\lambda$. The monodromy matrix from (\ref{dif1}) 
may be written as: \be T(x,y,\lambda) = {\cal P} exp \Big \{
\int_{y}^x {\mathbb U}(x',t,\lambda)dx' \Big \}. \label{trans} \ee
The fact that $T$ also satisfies equation (\ref{dif1}) will be
extensively used to get the relevant integrals
of motion. Compatibility conditions of the two differential
equations (\ref{dif1}), (\ref{dif2}) lead to the zero curvature
condition \cite{AKNS}--\cite{ZSh} \be \dot{{\mathbb U}} - {\mathbb V}' + \Big [{\mathbb
U},\ {\mathbb V} \Big ]=0, \label{zecu} \ee giving rise to the
corresponding classical equations of motion of the system under
consideration.

Hamiltonian formulation of the equations of motion is available again under
the $r$-matrix approach. In this picture the underlying
classical algebra is manifestly analogous to the quantum case.
The existence of the Poisson structure for ${\mathbb U}$ realized by the classical r-matrix,
satisfying the classical Yang-Baxter equation (\ref{clyb}), guarantees the integrability of
the classical system. Indeed assuming that the operator ${\mathbb U}$ satisfies the following
ultralocal form of Poisson brackets
\be \Big \{{\mathbb U}_a(x, \lambda),\ {\mathbb U}_b(y, \mu) \Big \} =
\Big [r_{ab}(\lambda - \mu),\ {\mathbb U}_a(x, \lambda) +{\mathbb U}_b (y,\mu) \Big ]\
\delta(x-y), \label{ff0} \ee
then $T(x,y,\lambda)$
satisfies (\ref{clalg}), and consequently one may readily show for a
system on the full line: \be \Big \{\ln tr\{T(x,y,\lambda_1)\},\ \ln
tr\{T(x,y, \lambda_2)\} \Big\}=0 \ee i.e. the system is
integrable, and the charges in involution --local integrals of
motion-- are obtained by expansion of the generating function $\ln
tr\{T(x,y,\lambda)\}$, based essentially on the fact that $T$
satisfies (\ref{dif1}).

Our aim here is to consider integrable models on the interval
with consistent ``boundary conditions'', and
derive rigorously the Lax pairs associated to the entailed
boundary local integrals of motion as a continuous extension of
the discrete case described previously. We
briefly describe this process below for any classical integrable
system on the interval. In this case one constructs a modified
transition matrix ${\cal T}$, based on Sklyanin's formulation and
satisfying again the Poisson bracket algebras ${\mathbb R}$ or
$~{\mathbb T}$. To
construct the generating function of the integrals of motion one
also needs $c$-number representations of the algebra ${\mathbb R}$
or ${\mathbb T}$ satisfying (\ref{refc}), (\ref{kso}).
The modified transition matrices, realizing the corresponding algebras
${\mathbb R},\ {\mathbb T}$ are given by (\ref{reps}), where now $T$ defined in (\ref{trans})
and ${\hat T}$ in (\ref{notation}).
The generating function of the involutive
quantities is defined in (\ref{gen}) and one shows in this case as well:
\be \Big \{t(x,y,t,\lambda_1),\
t(x,y,t,\lambda_2) \Big \} =0, ~~~ \lambda_1,\ \lambda_2 \in
{\mathbb C}. \label{bint0} \ee

In the case of open boundary conditions, exactly as in the discrete integrable models,
we prove (for more details on the proof see \cite{avandoikou})
\be \Big \{{\cal
T}_a(0, -L,\lambda),\ {\mathbb U}_b(x, \mu) \Big \} = {\mathbb M}_a'(x,
\lambda, \mu) + \Big [{\mathbb M}_a(x, \lambda, \mu),\ {\mathbb
U}_b(x,\ \mu ) \Big] \label{last} \ee where we define \be {\mathbb
M}(x,\lambda, \mu) &=& T(0, x, \lambda) r_{ab}(\lambda -\mu) T(x,
-L, \lambda) K^-(\lambda) \hat T(0, -L, \lambda) \non\\ &+& T(0,
-L, \lambda) K^-(\lambda) \hat T(x, -L, \lambda) \hat r^*_{ab}(\lambda
+\mu) \hat T(0, x, \lambda). \label{mm} \ee Finally bearing in mind the
definition of $t(\lambda)$ (\ref{gen}), and (\ref{last}) we conclude with: \be \Big
\{ \ln\ t(\lambda),\ {\mathbb U}(x, \mu) \Big \} = {\partial {\mathbb
V}(x,\lambda, \mu) \over \partial x} + \Big [ {\mathbb V}(x,\lambda, \mu),\
{\mathbb U}(x, \mu) \Big ]
\label{defin} \ee where \be {\mathbb V}(x,\lambda,
\mu) = t^{-1}(\lambda) \ tr_a \Big ( K^+(\lambda)\ {\mathbb M}_a(x , \lambda, \mu) \Big ).
\label{final1} \ee

As in the discrete case particular
attention should be paid to the boundary points $x =0,\ -L$. Indeed, for these
two points one has to simply take into account that $T(x,x
,\lambda) = \hat T(x,x, \lambda) ={\mathbb I}$. Moreover, the expressions derived in
(\ref{mm}), (\ref{final1})
are universal, that is independent of the choice of model.

\subsection{Example}

We shall now examine a particular example associated to the
rational $r$-matrix (\ref{rr}), that is the $gl_{{\cal N}}$ NLS
model. Although in \cite{paper} an extensive analysis for both
types of boundary conditions is presented, here we shall focus on
the simplest diagonal ($K^{\pm} ={\mathbb I}$) boundary
conditions. The Lax pair is given by the following
expressions \cite{ft, foku}: \be {\mathbb U} = {\mathbb U}_0 +
\lambda {\mathbb U}_1, ~~~{\mathbb V} = {\mathbb
V}_0+\lambda{\mathbb V}_1 +\lambda^2 {\mathbb V}_2 \label{lax0}\ee
where \be && {\mathbb U}_1 = {1\over 2i} (\sum_{i=1}^{{\cal
N}-1}E_{ii} -E_{{\cal N}{\cal N}}), ~~~~{\mathbb U}_0 =
\sum_{i=1}^{{\cal N}-1}(\bar \psi_i E_{i{\cal N}} +\psi_i E_{{\cal N}i}) \non\\
&&{\mathbb V}_0 = i \sum_{i,\ j=1}^{{\cal N}-1}(\bar \psi_i \psi_j E_{ij}
-|\psi_i|^2E_{{\cal N}{\cal N}}) -i\sum_{i=1}^{{\cal N}-1} (\bar \psi_i' E_{i{\cal N}} -
\psi_i' E_{{\cal N}i}), \non\\ && {\mathbb V}_1= -{\mathbb U}_{0},
~~~{\mathbb V}_2= -{\mathbb U}_{1} \label{lax} \ee and $\psi_i,\
\bar \psi_j$ satisfy\footnote{The Poisson structure for the
generalized NLS model is defined as: \be \Big \{ A,\ B  \Big \}= i
\sum_{i} \int_{-L}^{L} dx \Big ({\delta A \over \delta \psi_i(x)}\
{\delta B \over \delta \bar \psi_i(x)} - {\delta A \over \delta
\bar \psi_i(x)}\ {\delta B \over \delta \psi_i(x)}\Big )  \ee}:
\be \Big \{ \psi_{i}(x),\ \psi_j(y) \Big \} = \Big \{\bar
\psi_{i}(x),\ \bar \psi_j(y) \Big \} =0, ~~~~\Big \{\psi_{i}(x),\
\bar \psi_j(y) \Big \}= \delta_{ij}\ \delta(x-y). \ee

The boundary Hamiltonian for the generalized NLS model
may be expressed as (see \cite{paper}) \be {\cal H}&=& \int_{-L}^0 dx\
\sum_{i=1}^{{\cal N}-1}\Big (\kappa |\psi_{i}(x)|^2
\sum_{j=1}^{{\cal N}-1}|\psi_j(x)|^2 +\psi'_i(x) \bar \psi'_i(x)\Big )
\non\\ &-& \sum_{i=1}^{{\cal N}-1} \Big (\psi'_i(0) \bar \psi_i(0)
+ \psi_i(0) \bar \psi'_i(0)\Big ) + \sum_{i=1}^{{\cal N}-1}  \Big
(\psi_i'(-L) \bar \psi_i(-L) + \psi_i(-L) \bar \psi'_i(-L)\Big ).
\ee One sees here that the $K$-matrix indeed contributes as a genuine boundary effect.
The Hamiltonian, obtained as one of the charges in
involution (see e.g. \cite{paper} for further details) provides the
classical equations of motion by virtue of: \be && {\partial
\psi_i(x,t)\over
\partial t} = \Big \{{\cal H}(0,-L),\ \psi_i(x,t) \Big \}, ~~{\partial
\bar \psi_i(x,t)\over \partial t} = \Big \{{\cal H}(0,-L),\ \bar
\psi_i(x,t) \Big \}, \non\\ && -L \leq x \leq 0. \label{eqmo0} \ee
Indeed considering the Hamiltonian ${\cal H}$, we end up with the
following set of equations with Dirichlet type boundary conditions\be && i {\partial \psi_i(x,t) \over
\partial t} = - {\partial^2 \psi_i(x,t)\over  \partial^2 x} +2\kappa
\sum_{j=1}^{{\cal N}-1}|\psi_{j}(x,t)|^2\psi_i(x,t) \non\\ && \psi_i(0) = \psi_i(-L) =0 ~~~~~i
\in \{1, \ldots ,{\cal N}-1\}.  \label{eqmo2} \ee For a detailed and quite exhaustive analysis
of the various integrable boundary conditions of the NLS model see \cite{paper}.

As mentioned our ultimate goal here is to derive the boundary Lax
pair, in particular the ${\mathbb V}$ operator. Hereafter we shall focus on the SP
case with the simplest boundary conditions i.e. $K^{\pm} ={\mathbb
I}$. For any $gl_{\cal N}$
$r$-matrix we may expand
(\ref{final1}), taking also into account (\ref{rr}), in powers of
$\lambda^{-1}$ (we refer the interested reader to \cite{paper, avandoikou}
for technical details) and we obtain ${\mathbb
V}^{(3)}(x,\lambda)$ --the bulk part-- coincides with ${\mathbb
V}$ defined in (\ref{lax0}), (\ref{lax}), and for the boundary
points $x_b \in \{0,\ -L\}$ in particular: \be {\mathbb
V}^{(3)}(x_b, \lambda) = -{\lambda^2 \over 2i} \Big
(\sum_{i=1}^{{\cal N}-1} E_{ii} -E_{{\cal N}{\cal N}} \Big ) + i \sum_{i, j =1} ^{{\cal N}-1}
\bar \psi_i(x_b) \psi_j(x_b) E_{ij}  -i \sum_{i,j =1}^{{\cal N}-1} \Big (
\bar \psi'_i(x_b) E_{i{\cal N}} -\psi'_i(x_b) E_{{\cal N}i} \Big). \non\\ \ee We
may alternatively rewrite the latter formula as: \be {\mathbb
V}^{(3)}(x_b, \lambda) = {\mathbb V}(x_b, \lambda) +i
\sum_{i=1}^{{\cal N}-1} |\psi_i(x_b)|^2 E_{{\cal N}{\cal N}} + \lambda \sum_{i=1}^{{\cal N}-1}
(\bar \psi_i(x_b) E_{i{\cal N}} + \psi_i(x_b) E_{{\cal N}i}). \ee The last two
terms additional to ${\mathbb V}$ (\ref{lax0}), (\ref{lax}) are
due to the non-trivial boundary conditions; of course more
complicated boundary conditions would lead to more intricate
modifications of the Lax operator ${\mathbb V}$.

\noindent{\bf Acknowledgments:} We wish to thank the organizers of RAQIS'07 where this work was presented.
J.A. thanks INFN Bologna, and University of Bologna for hospitality.
This work was supported by INFN, Bologna section, through grant TO12.

\end{document}